\documentstyle[prl,aps,amsfonts,amssymb,twocolumn,epsfig]{revtex}
\begin{document}
\draft

\newcommand{\w}{\omega}
\newcommand{\kf}{k_F}
\newcommand{\G}{{\bf G}}
\newcommand{\A}{\text{\AA}}
\newcommand{\IM}{\text{Im}}
\newcommand{\RE}{\text{Re}}
\renewcommand{\vec}[1]{{\bf #1}}
\wideabs{
\title{Conductivity of a  clean one-dimensional wire}
\author{A. Rosch and N. Andrei}
\address{Center of Materials Theory, Rutgers University, Piscataway, NJ
08854-0849, USA}
\date{\today}
\maketitle

\begin{abstract}

We study the low-temperature low-frequency conductivity $\sigma$ of an
interacting one dimensional electron system in the presence of a
periodic potential.
The conductivity is strongly influenced by
conservation laws, which, we argue, need be violated
by at least {\em two} non-commuting Umklapp processes to
render  $\sigma$ finite.  The resulting dynamics of the slow modes is studied
 within a memory {\em matrix} approach, and we find exponential increase
as the temperature is lowered, $\sigma \sim (\Delta n)^2 e^{T_0/(N T)}$ 
close to commensurate filling $M/N$, $\Delta n=n-M/N \ll 1$, and 
$\sigma \sim e^{(T'_0/T)^{2/3}}$
 elsewhere.

\end{abstract}
\pacs{73.50.Bk,72.10.Bg,71.10.Pm}
}

The finite-temperature conductivity of a clean one-dimensional wire
\cite{review} is a fundamental and much studied question. Clearly the
``bulk'' conductivity of a wire in the {\em absence} of a periodic
potential is infinite even at finite temperatures $T$.  In this case
the conductance is independent of the length of the wire and is
determined by the contacts only.
Surprisingly, much less is known
about the conductivity in the presence of Um\-klapp scattering induced
by a periodic potential.  There is not even an agreement whether it is
finite or infinite at finite temperatures for generic systems
\cite{giamarchi,millis,zotos,zotos2,everybody}. We shall show that the
correct answer emerges when all relevant (weakly violated)
conservation laws are taken into account.  Those conservation laws are
exact at the Fermi surface and are violated by Umklapp terms away from
it.  We shall study the associated slow modes by means of a {\it
  memory matrix} formalism able to keep track of their dynamics. It
will allow us to calculate reliably the low temperature, low frequency
conductivity.

The topology of the Fermi surface of a $1d$ metal determines its
  low-energy excitations. Two well defined
 Fermi-points exist at momenta $k=\pm k_F$,  allowing us to define
 left and right moving
excitations, to be described by $\Psi_{L/R,\sigma=\uparrow 
\downarrow}$. We shall include in the fields 
 momentum modes extending to the edge of the Brillouin zone,
usually omitted in treatments that concentrate
on physics very close to the Fermi-surface.

The  Hamiltonian, including high energy processes, is 
\begin{eqnarray}
H=H_{LL}+H_{\text{irr}}+\sum_{n,m}^\infty H_{n,m}^U.
\end{eqnarray}
 $H_{LL}$ is the well-known Luttinger liquid Hamiltonian 
capturing the low energy behavior\cite{review},
\begin{eqnarray}
H_{LL}&=&v_F \int  \left(\Psi_{L
\sigma}^\dagger i \partial_x \Psi_{L \sigma}- \Psi_{R\sigma}^\dagger
i \partial_x \Psi_{R\sigma} \right)  + g \int  \rho(x)^2  \nonumber
\\ &=& \frac{1}{2} \int \frac{dx}{2 \pi}
\sum_{\nu=\sigma,\rho}  v_\nu\left( K_\nu (\partial_x
\theta_\nu)^2+\frac{1}{K_\nu} (\partial_x \phi_\nu)^2 \right) \nonumber
\end{eqnarray}
$v_F$ is the  Fermi velocity, $g>0$
 measures the strength of interactions, $\rho=\rho_L+\rho_R$
 is the sum of the left and right moving electron densities. 
In the  second line we wrote the
 bosonized\cite{review}
version of the Hamiltonian. Here 
 $v_\sigma$, $v_\rho$ are the spin and charge velocities, and  the
 interactions determine the Luttinger parameters $K_\nu$ with
 $v_\nu K_\nu=v_F$,
 $v_\rho/K_\rho=v_F+g/\pi $, $v_\sigma/K_\sigma=v_F-g/\pi$.

 The high energy processes are captured in the subsequent terms which
 are formally irrelevant at low energies (we
 consider only systems away from a Mott transition, i.e. away from half
 filling). Some of them, however, determine the low-frequency behavior
 of the conductivity at any finite $T$, since they induce the
 decay of the conserved modes of $H_{LL}$ (they are ``dangerously
 irrelevant'').  We classify these irrelevant terms with the help of
 two operators which will play the central role in our discussion. The
 first one is the translation operator $P_T$ of the right- and
 left-moving fields, the second one, $J_0 =N_R-N_L$, is the difference
 of the number of right- and left-moving electrons, and is up to
 $v_F$, the charge current of $H_{LL}$:
\begin{eqnarray}\label{PT}
P_T&=&\sum_\sigma \int dx \left(\Psi_{R
\sigma}^\dagger (-i \partial_x) \Psi_{R \sigma}+ \Psi_{L\sigma}^\dagger
(-i \partial_x) \Psi_{L\sigma}\right) \\ \label{J0}
J_0&=&N_R-N_L= \sum_{\sigma} \int dx  \left( \Psi_{R
\sigma}^\dagger \Psi_{R\sigma}-\Psi_{L\sigma}^\dagger
\Psi_{L\sigma} \right)
\end{eqnarray}
Both $P_T$ and $J_0$ are conserved by $H_{LL}$;  their
importance for transport properties is due to the
fact that both stay {\em  approximately} conserved 
in {\em any} one dimensional metal (away from half filling): 
processes which 
 change $J_0$ are forbidden {\it close to the Fermi surface} by
 momentum conservation. 
The linear combination
$P_0=P_T + k_F J_0$ can be identified with the total momentum
of the full Hamiltonian $H$ and is therefore also approximately
conserved.

We proceed to the classification of the formally irrelevant terms 
in the Hamiltonian. This classification  allows us to select all those 
terms (actually few in number)
that determine the current dynamics.
$H_{\text{irr}}$ includes all terms in $H-H_{LL}$
 which  commute with both $P_T$ and $J_0$, such as
 corrections due to the finite band curvature,
 due to finite-range interactions and similar terms. We will not need
their explicit form.

The Umklapp terms  $H^U_{n,m}$ ($n,m=0,1,...$) convert $n$ right-movers 
to left-movers (and vice versa) picking up lattice momentum
 $m 2 \pi/a=m G$, and do not commute 
with either $P_T$ or $J_0$. Leading terms are of the form,
\begin{eqnarray}\label{HU}
H^U_{0,m}&\approx&  g^U_{0,m} \int  e^{i \Delta k_{0,m} x} (\rho_L+\rho_R)^2 + h.c. \\
H^U_{1,m}&\approx&  g^U_{1,m} \sum_{\sigma}\int  e^{i \Delta k_{1,m} x} \Psi^\dagger_{ R \sigma} \Psi_{ L \sigma} 
\rho_{-\sigma}+h.c. \\
H^U_{2,m}&\approx&
 g^U_{2,m} \int  e^{i \Delta k_{2,m} x} \Psi^\dagger_{ R \uparrow}\Psi^\dagger_{ R \downarrow} 
\Psi_{ L \downarrow}\Psi_{ L \uparrow}+h.c. 
\end{eqnarray}
with momentum transfer $
\Delta k_{n,m}= n 2 k_F-m G$. A process 
transfering $n>1$ electrons 
  with total
spin $ n_s/2$ pointing in the  $z$-direction can be neatly expressed 
as
\begin{eqnarray} \label{boso}
H^U_{n,m}=   \frac{ g^U_{n,m,n_s}}{(2 \pi \alpha)^n} \int 
e^{i \Delta k_{n,m} x}
e^{ i  \sqrt{2} (n \phi_{\rho} +  n_s \phi_{\sigma}) }+ h.c.,
\end{eqnarray} 
$\alpha$ being a cut-off, of the order of the lattice spacing.
In fermionic variables the integrand
takes the form   
  $ \prod_{j=0}^{n/2-1} 
\Psi_{R \uparrow}^\dagger(x+j \alpha)
   \Psi_{R \downarrow}^\dagger(x+j \alpha)
  \Psi_{L \downarrow}(x+j \alpha)
  \Psi_{L \uparrow}(x+j \alpha)$ (for $n_s=0$ and even $n $).  

Note, though, that any {\em  single} term $H^U_{nm}$
conserves a linear combination of $J_0$ and $P_T$,
\begin{eqnarray}
\left[ H^U_{nm},\Delta k_{nm} J_0 + 2 n P_T \right]=0.
\end{eqnarray}
Indeed,  a term of the form (\ref{boso}) would appear
in  a continuum model
{\em without} Umklapp scattering, but with a Fermi momentum
$\tilde{k}_F=\Delta k_{nm}/(2n)$. In such a model, $ \Delta k_{nm} J_0/(2 n)
+ P_T$ is the total momentum of the system and 
therefore conserved. The importance of this simple but essential
conservation law has to our knowledge not been sufficiently realized
in previous calculations of the conductivity.  Due to this
conservation law a single Umklapp term can never induce a finite
conductivity! At least two independent Umklapp terms are required to
lead to a complete decay of the current. Further, two incommensurate
Umklapp terms
suffice to generate the rest.

To calculate the conductivity it is necessary to
keep track of the nearly conserved quantities and their
relation to the current. We will develop a
description of the slowest variables using the Mori-Zwanzig memory
functional \cite{forster,woelfle,giamarchi}.  Approximations
within this scheme  amount to short-time
expansions. In general, the short time decay of a quantity carries
little information on its long-time behavior; this,
however, is {\em not} the case for the slowest variables in the system,
where the short time and hydrodynamic behavior coincide. 


To set up the formalism \cite{forster} we define a scalar product
$(A|B)$ in the space of {\em operators}, 
\begin{eqnarray}
\left(A(t)|B\right)&\equiv& \frac{1}{\beta} \int_0^\beta d\lambda
\left\langle A(t)^\dagger B(i \lambda) \right\rangle, 
\end{eqnarray}
where we use the usual Heisenberg picture with $A(t)=e^{i H t} A e^{-i
H t}$. 
 We choose a  set ``slow'' operators  $j_1, j_2, ... j_N$
 which includes $j_1=J$, the full current operator. Standard arguments
\cite{forster} lead to the electric conductivity,
\begin{eqnarray} \label{sigmaE}
\sigma(\w,T)&=&\left[\left(\hat{M}(\w,T)- i \w \right)^{-1}
\hat{\chi}(T) \right]_{11}.  \end{eqnarray} 
Here $\hat{\chi}_{pq}=\beta (j_p|j_q)$ is the
matrix of the static $j_p j_q$ susceptibilities (as usually defined),
and $\hat{M}$
 is the matrix of memory functions  given by the projected
correlation functions of time-derivatives of 
the ``slow'' operators,
\begin{eqnarray}\label{M}
\hat{M}_{pq}(\w)=\beta \sum_r 
\left(\partial_t j_q \left| Q \frac{i}{\w-QLQ} Q 
\right| \partial_t j_r \right) (\hat{\chi}^{-1})_{rp}.
\end{eqnarray}
 The Liouville ``super''-operator, $L$, is defined by $L
A=[H,A]$ and $Q$ is the projection operator 
on the space perpendicular to the slowly varying variables $j_p$,
\begin{eqnarray}\label{Q}
Q=1-\sum_{pq} |j_q) \beta (\hat{\chi}^{-1})_{qp} (j_p|.
\end{eqnarray} 
We assumed for simplicity that all $j_p$ have the same signature under
time reversal.

The perturbative expansion of the memory matrix $\hat{M}$ is accompanied
by factors $1/\w$ guaranteeing it is always valid at short times. It
is also valid for small frequencies provided the slowly evolving
degrees of freedom are projected out (by the operator $Q$). Unlike the
 conductivity it is expected to be
 a smooth function of the coupling constants  which can  be
perturbatively evaluated.

We first consider a situation where some linear 
combinations of the $j_p$ are
conserved by $H$, in which case an infinite 
conductivity is expected. We introduce ${\cal P}_c$, the projection operator 
on the space of conserved currents, and carry out the required
matrix inversion  to find, 
\begin{eqnarray}\label{sigmajjj}
\sigma(\w \to 0,T>0)&=&
\sigma_{\text{reg}}(\w,T)
+i \frac{(\hat{\chi} \hat{\chi}_c^{-1} \hat{\chi})_{11}}{\w+i 0},
\end{eqnarray}
where $\hat{\chi}_c^{-1} = {\cal P}_c({\cal P}_c \hat{\chi} {\cal
  P}_c)^{-1}{\cal P}_c$.  Within any simple (short-time)
approximation, $\sigma_{\text{reg}}(\w,T)$ as defined above, is
regular (this approximation fails e.g. if some conserved current
$\hat{j}$ is not included in $j_1 ...j_N$).  Hence the Drude weight
$D(T)$ is finite at finite temperatures, $ \RE \sigma(\w \to 0)=2 \pi
D(T) \delta(\w)=\pi (\hat{\chi} \hat{\chi}_c^{-1} \hat{\chi})_{11}
\delta(\w)$. It is determined by the ``overlap'' of the physical
current operator $J$ with the conserved quantities $\chi_{1 s}$, $s$
labeling the conserved currents.  Remarkably, our perturbative
approximation is in accord with an exact inequality \cite{zotos2} for
the Drude weight, $D(T)\ge \frac{1}{2} (\hat{\chi} \hat{\chi}_c^{-1}
\hat{\chi})_{11}$.
Note that $\hat{\chi}$ can be calculated to an arbitrary degree
of precision around a Luttinger liquid and that the lower bound 
can be improved by including more conserved quantities 
\cite{zotos2}.

 Now consider the more realistic situation where the previously
conserved currents decay slowly (via Umklapp processes), 
in which case a finite conductivity
is expected.  We restrict ourselves to the two-dimensional space
spanned by $v_F J_0$ and $P_T$, which we argue have the longest decay
rate and dominate the transport. Here we approximate $J \approx v_F
J_0$ to keep the presentation simple. This affects only the high
frequency behavior of the conductivity \cite{millis}. There is a large
number of other nearly conserved quantities.  For example
$H_{LL}+H^U_{21}$, the relevant low-energy model close to half
filling, is integrable and therefore is characterized by an {\em
infinite} number of conservation laws. We can, however, neglect them
at low $T$ if our initial model is not integrable, expecting
that practically all conservation laws are destroyed by (formally
irrelevant) terms {\em close to the Fermi surface} leading to decay
rates proportional to some power of $T$. This is to be compared to
$J_0$ and $P_T$ which commute with {\em all} scattering processes at
the Fermi surface, leading to exponentially large lifetimes.

We now proceed to calculate the Memory matrix. To  leading order in the 
perturbations
 we can replace $L$ in (\ref{M}) by 
$L_{LL}=[H_{LL}, .]$ \cite{LL}, since
 $\partial_t v_F J_0$ and $\partial_t  P_T$
are already linear in $g^U_{n,m}$. As $L_{LL} P_T=L_{LL} J_0=0$, 
there is no contribution from the projection operator $Q$. The memory
matrix takes the form,
\begin{eqnarray}
\hat{M}&\approx&  \sum_{nm} M_{nm}(\w,T) \left(
\begin{array}{cc} 
v_F^2 (2 n)^2 & -2 n v_F \Delta k_{nm} \\
 -2 n v_F \Delta k_{nm} & (\Delta k_{nm})^2
\end{array}
\right) \hat{\chi}^{-1} \nonumber 
\end{eqnarray}
where,
\begin{eqnarray}
\hat{\chi}&\approx& \left(
\begin{array}{cc} 
2 v_F/\pi & 0 \\
0 & \frac{\pi T^2}{3} \left(\frac{1}{v_\rho^3}+\frac{1}{v_\sigma^3}\right)
\end{array}
\right) \\ 
 M_{nm}&\equiv& (g^U_{nm})^2 M_n(\Delta k_{n,m},\w)\equiv \frac{
\langle F;F \rangle^0_\w-\langle F;F \rangle^0_{\w=0}}{i \w}. \nonumber
\end{eqnarray}
Here  $F=[J_0,H^U_{nm}]/(2 n) $ (for simplicity we drop the indices $n,m$
on F), and $\langle F;F\rangle^0_{\w}$ is the
 retarded correlation function of $F$ calculated with respect to
 $H_{LL}$.

The memory function $M_2$ of the $4 k_F-Q$ process $H^U_{21}$ was
 calculated by Giamarchi \cite{giamarchi}, (not
 considering the matrix structure of $\hat{M}$ required by the 
conservation laws.) Higher Umklapps are considered in \cite{millis}. For $n_s=0$ and  even $n$ the memory function due to  the term
(\ref{boso}) can be analytically calculated,
\begin{eqnarray} \label{M2}
M_n(\Delta k,\w)=&&\frac{2 \sin 2 \pi K^n_\rho}{\pi^4 \alpha^{2n-2} v_\rho } 
\left[\frac{ 2 \pi \alpha T}{v_\rho} \right]^{4 K^n_\rho-2} \frac{1}{i \w}\times  \nonumber\\
\times [B(K^n_\rho-&&i S_+,1-2 K^n_\rho) B(K^n_\rho-i S_+,1-2 K^n_\rho)  \nonumber \\
 - B(K^n_\rho-&&i S_+^0,1-2 K^n_\rho) B(K^n_\rho-i S_+^0,1-2 K^n_\rho)]
 \nonumber\\
&&\hspace*{-1cm}\approx 
\frac{\alpha^{2-2n}}{ \pi^2 \Gamma^2(2K^n_\rho) v_\rho T} 
  \left( \frac{ \alpha \Delta k}{2} \right)^{4K_{\rho}^n-2} 
e^{-v_\rho \Delta k /(2 T)} 
\nonumber \end{eqnarray}
 where $K^n_\rho=(n/2)^2K_\rho$,
$B(x,y)=\Gamma(x) \Gamma(y)/\Gamma(x+y)$ and $S_{\pm}=(\w\pm v_\rho
\Delta k)/(4 \pi T)$, $S_{\pm}^0=S_{\pm}(\w=0)$. The last line is
valid for $\w=0$ and $T \ll v_\rho \Delta k$.

The origin of the exponential factor is as follows: 
processes involving momentum transfer $\Delta k$ are associated with initial
and final states of energies
$v | \Delta k|/2$, which are exponentially suppressed. If only charge
degrees of freedom are involved $v=v_\rho$, otherwise  
$v=\text{min}(v_\sigma,v_\rho)=v_\sigma$ \cite{LL}.
 For 
 $T \ll  v_{\sigma} \Delta k_{nm} $,  $n_s>0$ and $\w=0$,  we have,
\begin{eqnarray}\label{MlargeN}
M_n(\Delta k)\sim   \frac{\left(\alpha T/v_\rho\right)^{n^2 K_\rho-1}
\left(\alpha \Delta k \right)^{n_s^2 K_\sigma-2}}{
\Gamma^2(n_s^2 K_\sigma/2) v_\sigma^2 \alpha^{2n-3}} e^{-v_{\sigma} \Delta k/(2 T)},
\end{eqnarray}
while for  $T \gg  v_{\rho} \Delta k_{nm}$: $ M_{n}\sim 
T^{n^2 K_\rho+n_s^2 K_\sigma-3}$.

Using the above expressions with only one Umklapp term 
leads to a finite
Drude weight (cf eq(\ref{sigmajjj})),
\begin{eqnarray}\label{DT}
D(T)\approx \frac{ v_\rho K_\rho}{\pi} \frac{1}{1 +T^2 
\frac{2 \pi^2 n^2 K_\rho}{3 (v_\rho \Delta k_{nm})^2} 
\left(1+\frac{v_\rho^3}{v_\sigma^3}\right)}.
\end{eqnarray}
in accord with the observation that one process $H^U_{nm}$ is not sufficient
to degrade the current.

\begin{figure}[t]
  \centering
\epsfig{width=.8 \linewidth,file=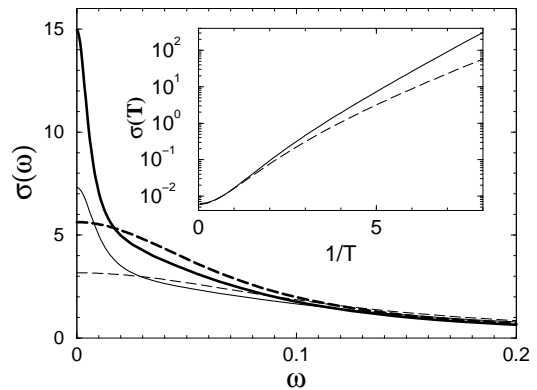}
\caption[]{The low frequency behavior of  $\sigma(\w)$ in the presence of two Umklapp terms for two different $T$. 
  The dashed lines are the result one obtains in conventional
  perturbation theory neglecting \cite{giamarchi} the matrix structure
  of $\hat{M}$ and the related conservation laws.  ($g_{20}=g_{21}=1$,
  $K_\rho=0.7$, $K_\sigma=1.3$, $\Delta k_{21}=-1.5 \Delta k_{20}$,
  thick lines $T=0.2$, thin lines $T=0.18$, $\w$ and $T$ measured in
  units of $v_\rho \Delta k_{20}$).  Note that two time scales appear
  - each describing the scale on which the associated conservation law
  is violated. The inset displays the $T$ dependence of
  $\sigma(\w=0)$.  } \label{fig2Umklapp}
\end{figure}

Only in the presence of a second incommensurate process $H^U_{n'm'}$
is the dc conductivity finite, 
\begin{eqnarray}\label{sigT}
\sigma(T, \omega=0)=\frac{(\Delta k_{nm})^2/M_{n'm'}+ (\Delta
k_{n'm'})^2/M_{nm}}{\pi^2 (n \Delta k_{n'm'}-n' \Delta k_{nm})^2}
\end{eqnarray} Note that the {\em slowest} process determines the
low-$T$ conductivity.  The
frequency and temperature dependence of the conductivity in the case
of two competing Umklapp terms is shown in Fig.~\ref{fig2Umklapp}.

The commensurate situation $\Delta k_{nm}=0$ requires extra
considerations. Whether the dominant scattering process $H_{nm}$ will
completely relax the current $J$ depends according to (\ref{sigmajjj})
on the overlap $\chi_{J P_T}$ ($[P_T,H_{nm}]=0$).  Using the
continuity equation for the charge, $\chi_{J P_T}$ can be related to
the deviation $\Delta \rho=2 \Delta n/a$ of the electron density
from commensurate filling with the remarkable identity
$\chi_{J P_T}=2 \Delta n/a+O(e^{-\beta\epsilon_F})$.  In a 3d lattice
of 1d wires, $ \Delta n$ is fixed by charge neutrality and is $T$
independent, in a single wire with contacts $ \Delta n$ varies at
low $T$ with $\Delta n(T) \sim T^2/(m v^3)$, where the mass $m$ is
a measure of the breaking of particle-hole symmetry, e.g. due to a
band-curvature $ k^2/2 m$.  In this case it is important to replace
$\Delta k_{nm}=0$ in Eqn. (\ref{DT}) or (\ref{sigT}) by $G \Delta n(T)$.

\begin{figure}[t]
  \centering
 \epsfig{width=.75 \linewidth,file=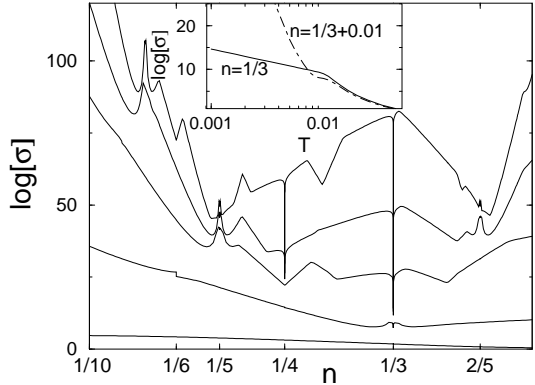}
\caption[]{Schematic plot of $\log[\sigma]$ as a function of the
  filling $n=2 \kf/G$ for various temperatures ($v G
  \beta=30,100,300,500,800$) based on an asymptotic
  (eqn.(\ref{MlargeN})) evaluation of (\ref{sigmaE}). Near
  commensurate fillings $n_c$ the conductivity is strongly
  enhanced at low temperatures but drops at $n=n_c$. The
  inset displays the $T$-dependence of $\sigma$ for $n=1/3$ and a
  filling very close to $1/3$ (dashed line). }
\label{fractal}
\end{figure}

Which of the various scattering processes will eventually dominate at
lowest $T$?  At intermediate temperatures, certainly low-order
(small $n$) scattering events win, being less suppressed by
 Pauli blocking.
At lower temperature  the exponential factors in
(\ref{MlargeN}) prevail and the processes with the smallest
$\Delta k_{nm}$ are favored.  We first analyze the situation close to
a commensurate point $k_F\approx G M_0/(2N_0)$. 
The {\it two} dominant processes are
 $H^U_{N_0 M_0}$
 with $\Delta k_{N_0 M_0}\approx 0$ and $H^U_{N_1 M_1}$ with $\Delta k_{N_1
M_1}=\pm G/ N_0$ (or $N_1 M_0 = \pm 1 \text{ mod } N_0$). The integer
 $N_1$  of  order  $N_0$, $N_1= \gamma_1 N_0$, depends strongly on
the precise values of $N_0$ and $M_0$. 
 We thus find that the d.c. conductivity at low $T$ is {\em largest}
close to commensurate points with, 
\begin{eqnarray}\label{sigmaKomm}
\sigma(k_F \approx  G M_0/(2 N_0))\sim (\Delta n(T))^2 \exp[\beta v G/(2 N_0)]
\end{eqnarray}
but $\sigma \sim T^{-N_0^2 K_\rho-(N_0 \text{ mod } 2)^2 K_\sigma+3}$ if the density 
is exactly commensurate with $|\Delta n(T)|< e^{-\beta G v/(4 N_0)}$.

To estimate the conductivity at a typical ``incommensurate'' point or
at commensurate points at temperatures not too low, we have to balance
algebraic and exponential suppression in (\ref{MlargeN}) by minimizing
$-\beta v G/(2 N) + (\gamma_1 N)^2 K \log[T]$ in a saddle-point
approximation to the sum over all Umklapp processes in ${\hat{M}}$. 
 Up to logarithmic
corrections we obtain $N_{\text{max}}^3 \sim \beta v G/( \gamma_1)^2$
and therefore for a ``typical'' incommensurate filling,
\begin{eqnarray}
\sigma_{\text{typical}}\sim \exp[c (\beta v G)^{2/3}]
\end{eqnarray}
where $c$ is a number depending logarithmically on $T$. At
present we cannot rule out that various logarithmic corrections sum up
to modify the power law in the exponent. We argue, however, that due
to the exponential increase (\ref{sigmaKomm}) of $\sigma$ at
commensurate fillings with exponents proportional to $1/N_0$, the
conductivity at small $T$ at any incommensurate point is smaller than
any exponential (but is larger than any power since any single process
is exponentially suppressed). In Fig.~{\ref{fractal}} we show
schematically the conductivity as a function of filling becoming more
and more ``fractal-like ''  for lower $T$.

Can the effects we predict be seen experimentally?  The complicated
structures as a function of filling shown in Fig.~\ref{fractal} are
 not observable in practice as they occur only at exponentially large
conductivities.  The $T$-dependence of the conductivity at
intermediate temperatures, however, 
 should be accessible, e.g. by comparing the
conductivities of clean wires of different length. Perhaps more
importantly, it is straightforward to apply our method to a large
number of other relevant situation, e.g. close to a Mott transition or
in the presence of 3d phonons, as we will discuss in a forthcoming
paper.

We thank R. Chitra, A.J. Millis, E. Orignac, A.E. Ruckenstein, S.
Sachdev, and P. W\"olfle for helpful discussions. Part of this work
was supported by the A.~v.~Humboldt Foundation and NSF grant
DMR9632294 (A.R.).


\begin{references}
\bibitem{LL} For large $\Delta k_{nm}$  corrections from
$H_{\text{irr}}$ like band-curvature terms are important, we neglect
them here
Their  inclusion leads again to
 exponential suppression in eq(\ref{MlargeN}) with modified numerical factors.
\end{references}
\end{document}